\documentclass[10pt,letterpaper]{article}
\usepackage{opex3}

\usepackage{hyperref}

\usepackage[T1]{fontenc}
\usepackage[cp1250]{inputenc}
\usepackage{graphicx}
\usepackage{amsmath}
\usepackage{cite}
\makeatletter
\newcommand{\rb}{$^{87}$Rb~}
\newcommand{\rbf}{$^{85}$Rb~}

\newcommand{\avg}[1]{\langle #1\rangle}

\renewcommand{\vec}[1]{\mathbf{#1}}
\newcommand{\appropto}{\mathrel{\vcenter{
\offinterlineskip\halign{\hfil$##$\cr
\propto\cr\noalign{\kern2pt}\sim\cr\noalign{\kern-2pt}}}}}
\makeatother
\begin{document}
\title{ Generation and delayed retrieval of spatially multimode Raman scattering in warm rubidium vapors}
\author{Rados{\l}aw Chrapkiewicz$^*$ and Wojciech Wasilewski}
\address{Faculty of Physics, University of Warsaw, ul. Ho\.{z}a 69, PL-00-681 Warszawa, Poland}
\email{$^*$radekch@fuw.edu.pl}
\begin{abstract}
We apply collective Raman scattering to create, store and retrieve spatially multimode light in warm rubidium-87 vapors. The light is created in a spontaneous Stokes scattering process. This is accompanied by the creation of counterpart collective excitations in the atomic ensemble -- the spin waves. After a certain storage time we coherently convert the spin waves into the light in deterministic anti-Stokes scattering. The whole process can be regarded as a delayed four-wave mixing which produces pairs of   correlated, delayed random images. 
Storage of higher order spatial modes up to microseconds is possible owing to usage of a buffer gas. 
We study the performance of the Raman scattering, storage and retrieval of collective excitations focusing on spatial effects and the influence of decoherence caused by diffusion of rubidium atoms in different buffer gases. 
We quantify the number of modes created and retrieved by analyzing   statistical correlations of intensity fluctuations between portions of the light scattered in the far field. 
\end{abstract}

\section{Introduction}
Room temperature atomic vapors are a popular medium for quantum information processing. Perhaps, they offer the simplest approach to quantum repeaters \cite{Duan2001b}, quantum memory \cite{Appel2008,Honda2008} and numerous other quantum-enhanced protocols possible with deterministic light-atom interface \cite{Hammerer2010} and long-lived ground state coherence. Room temperature vapors can be effectively utilized in two distinct ways. Atoms can be placed in a paraffin coated cell without a buffer gas which they traverse multiple times during the interaction so that their quantum state is symmetric with respect to the permutations and within the Holstein-Primakoff approximation the cell holds a single bosonic mode \cite{Jensen2010}. At the opposite extreme, the atoms are kept in a buffer gas which should virtually immobilize them. Then transverse and longitudinal patterns can be stored and interfaced to \cite{Nunn2008b,Vasilyev2010,Glorieux2012,Higginbottom2012}. 
  Room temperature vapors can also be a very effective four-wave mixing medium \cite{Boyer2008} and a source of spatially multimode squeezing \cite{Boyer2008b} with a potential for quantum imaging \cite{Brida2010}.
  
In this paper we characterize a multimode generator of  temporally separated random image pairs with a Raman interface. The images are formed at the write-in and the read-out of atomic excitations, accomplished by driving a spontaneous Stokes and a stimulated anti-Stokes scattering respectively. 
At the write stage we create a random scattered Stokes field, which forms a speckle pattern in the far field. It results from interference of distinct spatial modes. At the same time  coupled distribution of excitations arises within the atomic sample  \cite{Raymer1981,Koodynski2012}. This can be thought as a storage of the random but known image in the atomic memory. After a certain time the excitations can be converted to the anti-Stokes scattered light. 

In the experiments we use rubidium-87 vapor mixed with a buffer gas. 
We measure the effects of diffusion and deterioration of the effective optical depth due to the collisional broadening. 
  We quantify the spatial multimode characteristics and a capacity of the generator. 
In general this task is accomplished either by measuring field coherence \cite{Ribeiro1994} or by measuring second order Hanbury-Brown Twiss type correlations \cite{Marino2008,Boyer2008b}.
We have chosen to measure the latter, i.e. statistical correlation between intensity fluctuations in various scattering directions from large sets of single-shot images similarly as in \cite{Ferri2005}. We show that the results are consistent with a simple theoretical model \cite{Koodynski2012}.

In this particular operating regime we produce   multiple correlated modes of radiation that are released at separate time instants, first in the Stokes scattering process, second in the anti--Stokes scattering \cite{Koodynski2012}.
This is similar to recent work demonstrating entangled images \cite{Boyer2008}, however in our generator the  second image can be held in memory until the first image is processed or a suitable moment comes. In contrast to previous work \cite{Marino2009} it is achieved   without resorting to additional, separate quantum memory implemented in another atomic ensemble. 

In this article we examine purely classical properties of the correlated images. We focus on spatial properties of light and understanding atomic decoherence.
However, we note that our source shares a number of features with sources of non-classical light based on atomic ensembles \cite{ChouPRL04,Boyer2008}.
Our main motivation for the present study was that of possible future applications in multimode quantum information processing \cite{Lassen2007}, particularly involving spatially multimode squeezing which has attracted large attention recently \cite{Wagner2008,Janousek2009,Chalopin2011}.
We also suppose that our results might be of significance for quantum imaging { \cite{Boyer2008b,Brambilla2008,Brida2010}}, especially in applications which require storage of ghost images \cite{Cho2012}.
\section{Generator} \label{sec:model}
\begin{figure}
\includegraphics[width=\textwidth]{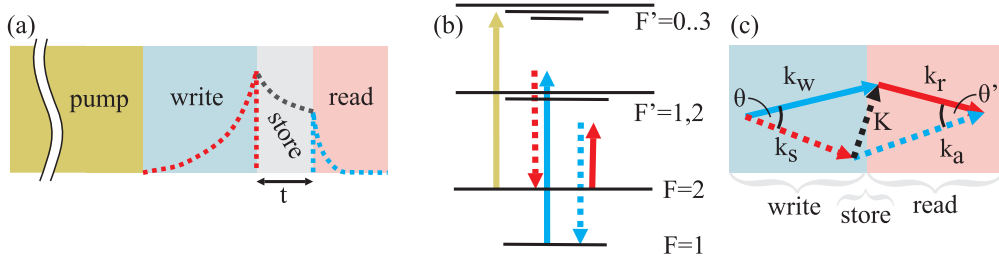}
\centering
\caption{The process of the write-in and the read-out of atomic spin waves. 
(a) Left panel illustrates the pulse sequence.
Dashed contours give a typical temporal evolution of the intensity of the Stokes and anti-Stokes scattered light as well as the decay of the atomic spin wave.
(b) As sketched in the middle panel, the pump pulse is resonant with the F=2$\,\rightarrow\,$F'=2 transition on D2 line, while write and read pulse are $1\,$GHz detuned from F'=1$\,\rightarrow\,$F'=1 and F=2$\,\rightarrow\,$F'=2 transitions on D1 line respectively.
(c) Right panel depicts the relation between the wave vectors of pumps $\vec k_w$, $\vec k_r$, photons $\vec k_s$, $\vec k_a$ and a spin wave $\vec K$ involved in the cycle.}
\label{fig:general}
\end{figure}
The operating cycle of the multimode delayed   images generator and the optical transitions are illustrated in Fig.~\ref{fig:general}.
The cylindrical cell containing rubidium-87 vapor in a buffer gas is initialized by optical pumping of the atoms into F=1 states of the ground manifold. Next, a pencil-shaped region is illuminated with a strong write beam. This induces scattering of anti-Stokes photons and creation of a spin wave corresponding to coherent excitation of atoms into F=2 state. The scattering grows exponentially due to the boson nature of the fields involved. We stop it when tens of thousands of excitations have been produced. After that we wait for an adjustable amount of time and illuminate the atomic sample with the read beam. It is tuned  in order to cause scattering of Stokes photons at the expense of the atomic excitations. It can be thought that the read beam probes the atomic coherence by converting it to the light.

In the write process the difference of wave vectors between the annihilated write laser photon and the created Stokes one $\vec k_w-\vec k_s$ is mapped onto the spatial phase of an atomic excitation. The excitation takes on a form of a coherent spin wave and acquires a wave vector $\vec K=\vec k_p-\vec k_s$. At the readout stage the stored phase given by the wave vector $\vec K$ adds to the read beam field wave vector $\vec k_r$ leading to scattering of anti-Stokes photons $\vec k_a$ in the direction opposite to the Stokes ones, as illustrated in Fig.~\ref{fig:general}c. 
With $z$-axis along the write beam and $\vec k_w$-vector and the scattering in the $x$--$z$ plane the spin wave wave vector $\vec K$ equals:
\begin{equation}
\vec K=\vec k_w-\vec k_s= \left[|k_s|\sin\theta,0,|k_s|\cos\theta -|k_w|\right]
\simeq 2\pi [\theta/(0.8\,\text{mm}\,\text{mrad}),0,1/(44\,\text{mm})]
\end{equation}
where $\theta$ is the scattering angle and the numbers correspond to 6.8~GHz Stokes shift (F=1$\,\rightarrow\,$F=2 scattering) at D1 line. 

Suppose the active part of the atomic sample has a pencil shape of a diameter $w$, set by the write and the read beam, and the length $L$ along $z$-axis. Within this volume atomic excitations which wave vectors $\vec K$ and $\vec K'$ differ more than the inverse dimensions of the active region: $|\vec K-\vec K'|>2\pi [1/w,1/w,1/L]$, are distinguishable and independent. This sets the fundamental limit on the number of independent modes which may be simultaneously stored in the atomic memory.
Since the scattered light couples to the respective spin wave $\vec K$-vectors the distinguishability of various $\vec K$ components of the atomic excitation also imprints on the scattered light. Conversely, this implies the emission into close directions has to be correlated   i.e. the light in the far field exhibits a finite coherence area.

In addition the Stokes scattering process populates only certain superpositions of the distinguishable   spatial modes. 
One of the easiest way to analyze it is to decompose the atomic excitations and likewise scattered light into a pairwise coupled basis, in which each light mode is coupled with exactly one atomic mode. 
The obtained light modes are all centered around the write beam $\vec k_w$ while the spin wave modes around the origin of the $\vec K$-vector space.
The modes resemble Laguerre-Gaussian functions \cite{Koodynski2012}. Importantly, the spatial shape of the modes is time independent. 
The width of the fundamental spin wave mode is approximately $\Delta K=\sqrt{|k_s|/L}$. 
The associated fundamental light mode is a Gaussian beam with a confocal parameter close to the sample length $L$. 
The Raman growth parameters of higher order modes slowly diminish along with mode number. 
Typically the number of significantly excited modes is close to the Fresnel number of the write beam $F=|k_w|w^2/(2\pi L)$ if the decoherence is neglected   \cite{Raymer1981}.
During each single write pulse Stokes modes are populated. They produce a speckle-like diffraction pattern in the far field.
  
Note that each speckle is not a single mode but an interference of many modes excited with random amplitudes and phases. 
Thus, every speckle has a different position, size and amplitude. 
Yet, an approximate average size of the speckle can still be worked out from the mode picture. 
Given $N$ similarly populated Laguerre-Gaussian modes we typically find single spots in their interference pattern to occupy about 
$1/\sqrt{N}$ of the solid angle of the fundamental mode, because the central lobe of the $n$-th mode is $\sqrt{n}$ times thinner than the extent of the fundamental mode. This is also equal to the coherence area of the image.
Conversely, the solid angle of the entire pattern is $N$ times bigger than the solid angle of a single coherence area assuming $N$ modes are populated. This can be justified twofold. Firstly $N$ modes interfere to $N$ independent coherence areas with uncorrelated intensities. Secondly the size of the pattern is determined by the highest order mode, which has $\sqrt{N}$ times bigger solid angle than the fundamental mode which in turn is $\sqrt{N}$ times bigger than the coherence area.
Note, that the above are approximate statements as the population of consecutive modes is diminishing slowly and the threshold above which the modes are counted as populated is vague. Therefore the number $N$ is approximate. 

In the experiment, the number of excited modes is limited by the diffusion. 
This is because high order modes are composed of components with large $\vec K$ which correspond to abrupt spatial variations. 
They are easily washed off by diffusion of the atoms in a buffer gas during the storage. 
Therefore in the experiment we expect the number of   retrieved modes $N$ to be lower than the Fresnel number of the write beam $F$. 

In our generator we observe an image created by Stokes photons and anti-Stokes photons in the far field, $n_s(\theta_x,\theta_y)$ and $n_a(\theta_x,\theta_y)$ respectively. 
In the ideal case of a plane-wave write and read beam we would expect that every point in $n_{s,a}(\theta_x,\theta_y)$ should carry information about the spin wave with $\vec K_\perp=|k_s|(\theta_x,\theta_y)$. In case of a finite write and read beam widths $w$ we expect the ideal images to be blurred due to beams wave vector spread. 
The true images should be close to the convolution of atomic excitations distribution in the sample with the read beam distribution in $\vec k_\perp$-space. 
\section{Experimental setup} \label{sec:setup}
\begin{figure}
\centering
\includegraphics[width=\textwidth]{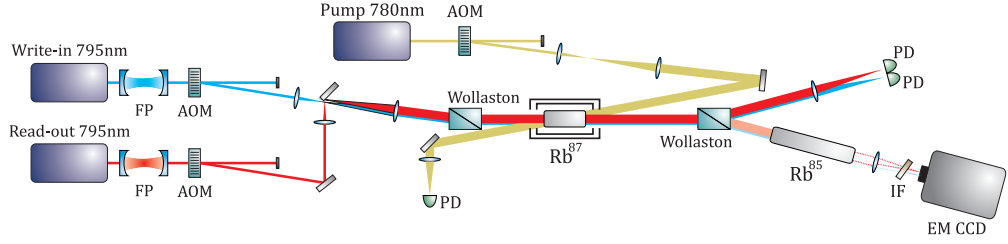}
\caption{The experimental setup. Write and read lasers were spectrally and spatially cleaned with Fabry--Perrot filters FP. The pulses were shaped by acousto-optic modulators (AOM). The shielded cell with \rb vapor was illuminated by the write, the read and the pump beam. The far field of the Raman scattering was observed on the electron multiplier CCD (EM CCD) camera and the background of the scattering beams was supressed by polarization and the filtering (Wollaston polarizer and the \rbf cell). Interference filter IF filtered the spontaneous emission background during the pumping. The pulse sequence was monitored using photodiodes PD.}
\label{fig:setup}
\end{figure}
The experimental setup is schematically presented in Fig.~\ref{fig:setup}. The pump laser (distributed feedback diode laser, Toptica) was resonant with the $\textrm{F}=2\rightarrow \textrm{F}'=2$ transition on the D2 line to pump the \rb atoms to $F=1$ ground level. 
Both the write and the read lasers were external cavity diode lasers from Toptica. They were detuned from the $\textrm{F}=1\rightarrow \textrm{F}'=1$ transition by 1 GHz to the red and by 1 GHz from the $\textrm{F}=2\rightarrow \textrm{F}'=2$ transition to the blue respectively. Both were additionally cleaned up spectrally as well as spatially by Fabry-Perrot cavities with FSR=13.6~GHz. 

The pulse sequence shown in Fig.~\ref{fig:general} was produced by acousto-optical modulators. 
The pump pulse duration 780~$\mu$s and its 4.2 mm full width at $1/e^2$ waist diameter ensured bleaching of the medium in the volume were the Raman scattering took place. 
The write pulse started immediately after the pump pulse. 
The $1/e^2$ beam diameter was 2.16~mm. 
Pulse duration varied from 200~ns to 10~$\mu$s and the power up to 17 mW. 
After the write pulse we inserted a variable storage delay. 
Finally a read pulse was directed onto the cell and converted the spin waves into anti-Stokes photons. The $1/e^2$ beam diameter was 1.76~mm.

The read beam was tilted by 13~mrad with respect to the write beam in order to direct the scattered photons onto separate regions on the camera. 
All beams crossed inside a cylindrical quartz cell (Precision Glassblowing), 100 mm long and 25 mm in diameter, which contained isotopically pure \rb. The cell was mounted inside a two-layer magnetic shielding and was heated with bifilar wound copper coils. For comparison we used three cells containing different buffer gas, 1 torr or 0.5 torr Krypton or 5 torr Neon.

The scattered light was separated from the write and read beams on a Wollaston polarizer. The residual laser background was then absorbed in the hot \rbf cell.   We decided to use this method due to its simplicity but one may alternatively use Fabry-Perrot interferometers \cite{Manz2007}. An interference filter was used to dump the 780~nm fluorescence emitted during the initial pumping. 
We checked that the laser background was of the order of a few photons per sequence and it was negligible compared to the residual fluorescence background and the camera noise.

We focused the far field of the Raman light onto an electron multiplier CCD camera (ImageEM, Hamamatsu).
We calibrated the position on the camera with respect to the beam propagation angle inside the cell. The pixel pitch corresponded to 76 $\mu$rad angle inside the cell. We also calibrated sensitivity with a power meter. Since the overlapping write and read beams were tilted with respect to each another, the corresponding Stokes and the anti-Stokes scattering light fell on distinct regions of the sensor. The camera was incapable of resolving the light temporally and we always captured both the Stokes and the anti-Stokes scattering integrated over the duration of write and read pulse. 
For each setting of the pulse lengths, powers and the storage time we captured 500 or 10000 separate images of the scattered light with a cycle rate of about 90 Hz. Despite the usage of an  interference filter before the camera we had some background coming from spontaneous emission in the pumping process. Thus the reference set of 100 frames with pump and write beams was saved for background subtraction.

We find that 500 frames are sufficient to calculate the average intensity of the scattered Stokes and anti-Stokes light $\avg{n_s(\theta_x,\theta_y)}$ and $\avg{n_a(\theta_x,\theta_y)}$ respectively. The average intensities presented below were calculated subtracting the background. The main origin of the background was the stray fluorescence of the atoms emitted during pumping. It was sufficiently stable to calculate it with high precision from 100 frames.
We find that the average intensity of the scattered light was symmetric around the driving beam. Therefore we averaged it with respect to the polar angle to obtain $\avg{n_s(\theta)}$ and $\avg{n_a(\theta)}$ with better signal to noise ratio. 
For calculating second order moments, including intensity correlations, we collected data sets containing 10000 frames.

Atom density in the main \rb cell was estimated by measuring the absorption spectrum for the write laser attenuated to a fraction of saturation intensity right before the cell. The result was fitted with a theoretical model to estimate the gas temperature and absorption at resonance. Cell heater temperature, measured by the resistance of the windings, was adjusted so as to obtain a Doppler-broadened optical depth of 130. 
The stability of the setup was verified by repeating this measurement prior to and right after the proper experimental sequence.
Note that the Doppler broadened optical thickness is not sensitive to the pressure broadening of the order of few natural linewidths. It is deemed that using the above method we obtain almost the same rubidium vapor pressure in all of the cells. 
\section{Results} \label{sec:results}
\paragraph{Decay of the spin waves}
Firstly we measured the decay rate of the spin waves vs. magnitude of their wave vector $|\vec K|$. It can be extracted from measurements of the average intensity of anti-Stokes $\avg{n_a(\theta_x,\theta_y)}$ as a function of the storage time for constant write pulse parameters.

To produce a widespread, significant excitation of spin waves in the cell we used short write pulse with a maximum available power of 17.5~mW. 
The pulse duration was adjusted to produce comparable spin wave excitations for each buffer gas resulting in $t_w=330\,$ns for 0.5 torr Krypton, $t_w=500\,$ns for 1 torr Krypton and $t_w=1.8\,\mu$s for 5 torr Neon. After a variable storage time $t_s$ a $2\,\mu$s read pulse is sent down the cell to convert remaining atomic excitation to anti-Stokes light. The average intensity of anti-Stokes $\avg{n_a(\theta_x,\theta_y)}$ is a Gaussian function of angles $\theta_x,\theta_y$. With increasing the storage time $t_s$ the height and width of this Gaussian decreases.

\begin{figure}
\centering\includegraphics[width=1\textwidth]{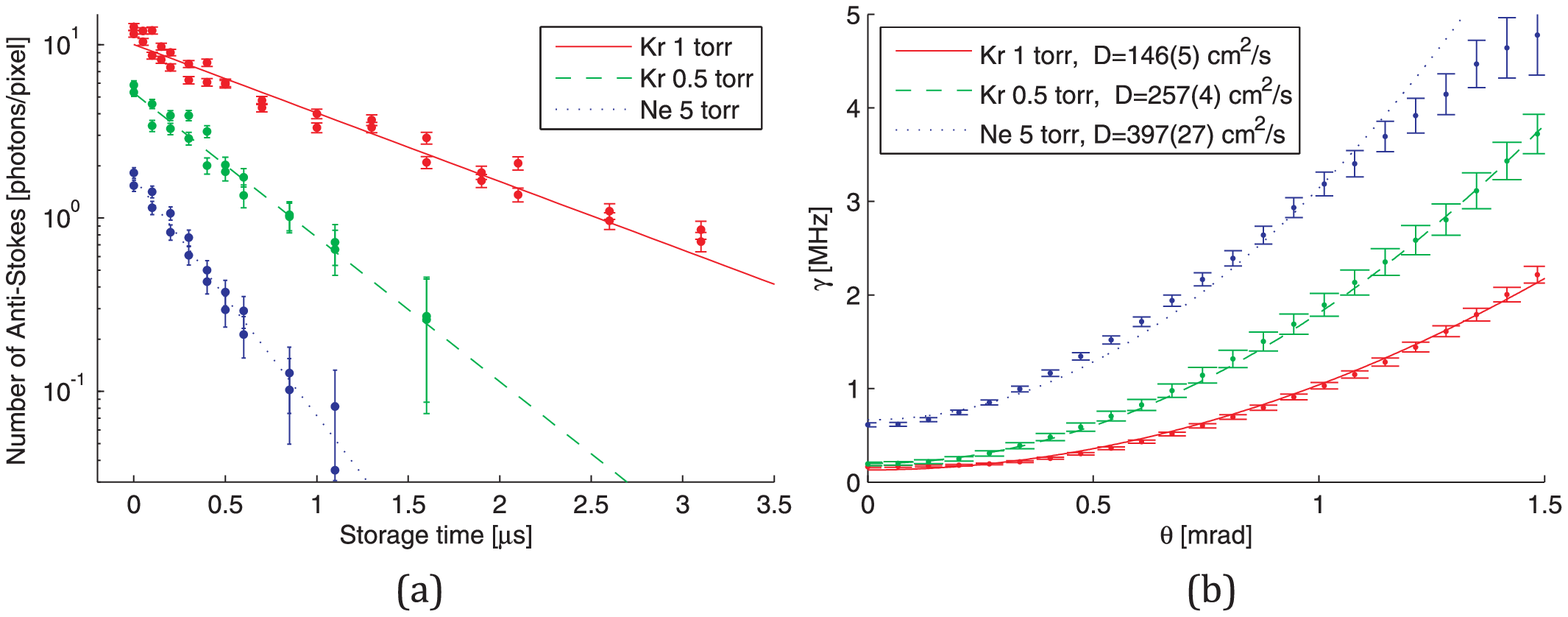}
\caption{(a) A typical decay of the average anti-Stokes intensity $\avg{n_a(\theta=1\,\text{mrad})}$ as a function of the storage time $t_s$ together with exponential fits $\exp(-\gamma t)$. 
(b) The decay rates $\gamma$ as a function of the scattering angle $\theta$ with a quadratic fit $\gamma=Dk_s^2\theta^2+\text{const}$. 
Results are presented for three different buffer gases.}
\label{fig:decays}
\end{figure}

Each direction of observation $(\theta_x,\theta_y)$ with respect to the read beam corresponds to a certain transverse anti-Stokes photon wave vector $\vec k_{a,\perp}=2\pi (\theta_x,\theta_y)/(0.8 \text{ mm mrad})$. 
It is approximately equal to the coupled spin wave wave vector $\vec K$ to within the uncertainty given by the read beam wave vector spread corresponding to the angular spread of $0.26\,$mrad. 
Therefore the evolution of the anti-Stokes intensity $n_a(\theta_x,\theta_y)$ as a function of storage time $t_s$ provides indication of the spin wave decay for the best coupled spin wave wave vector. In Fig.~\ref{fig:decays}(a) we present typical decays along with fitted exponentials. In Fig.~\ref{fig:decays}(b) we gathered decay rates as a function of angle between the read beam and scattered photons. All the errorbars plotted on the graphs correspond to the 1$\sigma$ uncertainty calculated from the variance of the experimental data. 

Similar to \cite{Glorieux2012} we attribute the decay of the spin wave to the diffusion of the rubidium atoms in the buffer gas.
The evolution of the atomic excitation is given by the diffusion equation, which leads to damping of spin waves. In particular the spin wave with wave vector $K$ decays with rate proportional to square of its wave vector $\gamma_K=DK^2$, with $D$ being the diffusion coefficient. The parabolic fits of Fig.~\ref{fig:decays}(b) give the diffusion rates equal 397(27) cm$^2$/s, 257(4) cm$^2$/s, 146(5) cm$^2$/s for 5 torr Neon, 0.5 torr Krypton and 1 torr Krypton respectively. The values measured for Krypton are consistent with collisions cross sections \cite{Gibble1991}. 
However, the value for Neon is higher than recently reported, by a factor of 4 \cite{Glorieux2012} or 40 \cite{Firstenberg2009}. 
We have no good explanation for this observation, however owing to large discrepancies between reported values we display the data as a reference for possible future studies.

\begin{figure}
\centering
\includegraphics[width=1\textwidth]{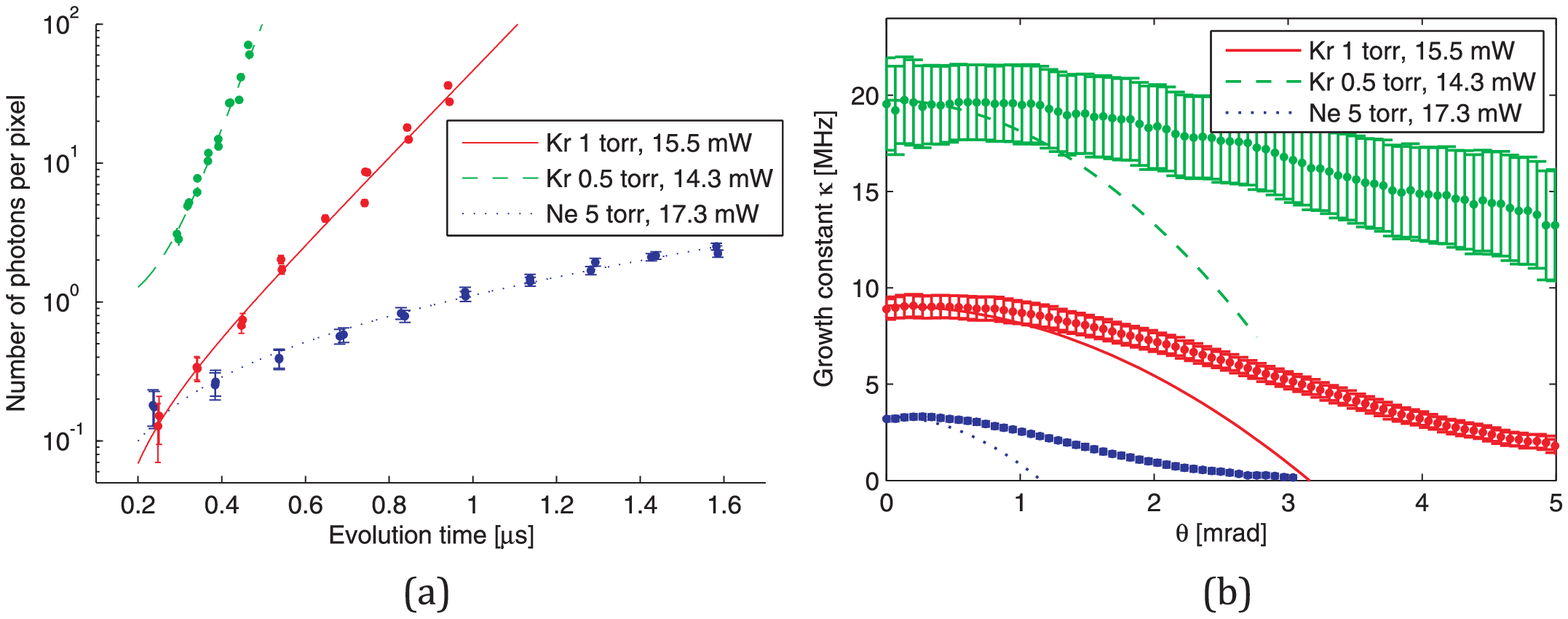}
\caption{(a) Typical growth of the average intensity of Stokes scattering $n_s(\theta)$ as a function of the write pulse duration $t_w$ for the scattering angle $\theta=2$~mrad with the exponential fit $\exp(\kappa t)+$const. The data taken for three different buffer gasses with similar atomic density. 
(b) Growth rate $\kappa$ as a function of the scattering angle (points with errorbars). The lines show the expected damping due to diffusion.}
\label{fig:growth}
\end{figure}

Let us note that alternatively the effect of diffusion can be described as Dicke-narrowed two-photon resonance, as explained in the context of EIT in \cite{Shuker2007,Firstenberg2012}. 
\paragraph{Spontaneous scattering and growth of the spin wave}
Knowing the intrinsic decay of the spin wave in the dark, we are in a position to study its growth during the write stage. Within the experimental precision the growth $n_s(\theta_x,\theta_y)$ as a function of write pulse length $t_w$ can be approximated as a simple exponential, as shown for typical situations in Fig.~\ref{fig:growth}(a). The duration of the write pulse is limited so that all saturation effects can be neglected. 

In the limit of plane wave driving beam, when each spin wave evolves separately, we would expect the growth rate $\kappa$ to be a sum of three components $\kappa=\zeta^2-\gamma_D-\gamma_\text{sp}$.
$\zeta^2$ denotes the coherent scattering rate which is directly proportional to the intensity of the pumping light $I_w$ and atomic density. It
drops for large scattering angles due to a reduced beam overlap between light fields involved. 
$\gamma_D$ denotes the diffusion-induced spin wave decay.
Finally, $\gamma_\text{sp}$ denotes the rate of the single atom spontaneous emission which depends linearly on the light power $I_w$ and has a contribution from the collisions with the buffer gas --- pressure broadening of the excited state. As a single atom effect $\gamma_\text{sp}$ should be independent from spin wave periodicity i.e. $|K|$.
In Fig.~\ref{fig:growth}(b) we present measured growth rate $\kappa$ as a function of angle $\theta$ between the write beam and Stokes photon. We also mark the predicted contribution of the diffusion-induced spin wave decay $\gamma_D$ inferred from the previous measurement.

As apparent in the figure, measured growth rate for high angles is faster than expected from the plane wave model. 
This is explained by the fact, that the fundamental mode of atomic excitation is not a plane wave. It has a large wave vector spread, approximately $\sqrt{k_s/L}$ which corresponds to an angular spread of 2.3 mrad. Fast coherent uprise of the fundamental mode boosts the growth rate at high angles.

We also note significant overall difference in scattering rate between various buffer gases despite comparable density of the atomic sample and write laser power. It can be attributed to the perturbation of the excited state by the collisions, studied in the case of Neon \cite{Manz2007}. 
The growth is slower at a higher pressure due to higher $\gamma_\text{sp}$ which results from the loss of phase coherence of the excited state.
As seen in Fig.~\ref{fig:decays}(b) and Fig.~\ref{fig:growth}(b) for 0.5 an 1 Torr Krypton one can trade slower diffusion for faster growth.
\paragraph{Intensity correlations}
\begin{figure}
\centering
\includegraphics[width=\textwidth]{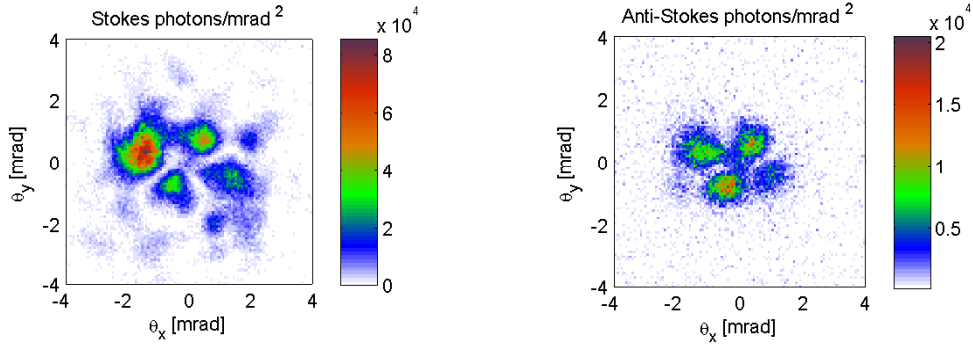}
\caption{The single shot, speckle-like picture of the Raman Stokes (left) and the anti-Stokes (right) scattering.   A single speckle comes from an interference between a larger number of spatial modes.
Note the pictures display similar features but are rotated by 180$^\circ$. { The anti-Stokes scattering is cropped for higher angles due to decoherence.} }
\label{fig:single}
\end{figure}
Let us now turn our attention to the   multimode spatial nature of scattered fields.
A typical single shot image captured by the camera is presented in Fig.~\ref{fig:single}. 
It was taken with the 1 torr Krypton as a buffer gas. 
The write and read pulse power and duration were 17.5~mW, 800~ns and 5~mW, 2~$\mu$s respectively. 
The frame shows the speckle pattern of both the Stokes and the anti-Stokes scattering with the highest intensity around the driving beam directions. The pattern is produced due to the boson amplification of initially random fields   in many spatial modes, just as it happens in spontaneous parametric down-conversion \cite{Jost1998} or in Bose-Einsten condensate collisions \cite{Perrin2007}. 

Looking carefully one can see that the anti-Stokes picture is similar to the Stokes rotated by $180^\circ$ around the center, although features at high angles are cropped. 

\begin{figure}
\centering
\includegraphics[width=\textwidth]{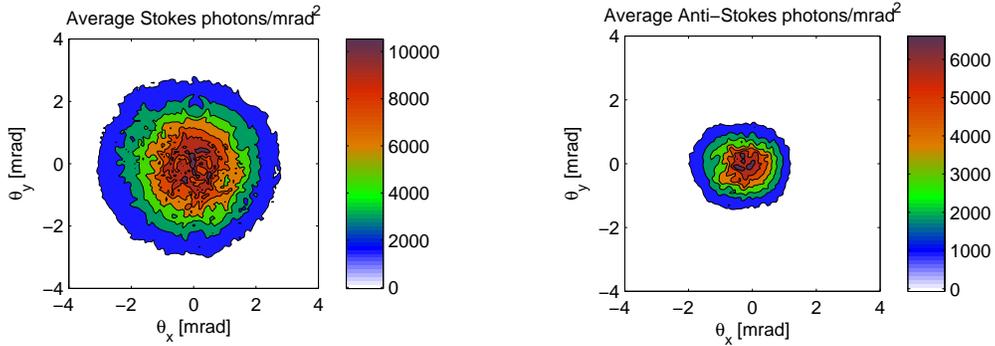}
\caption{The average of the 10000 of frames shows that the scattering light is emitted symterically around the direction of the scattering beams (the origin of the coordinates). Data collected with cell containing 1 torr Kr.}
\label{fig:avg}
\end{figure}

Upon averaging 10000 frames the speckle pattern disappears giving way to smooth, symmetric Gaussian intensity distributions as illustrated in Fig.~\ref{fig:avg}. This is because the spot position in each single shot is random. 
Again, the angular spread of the anti-Stokes is significantly smaller.
This is a consequence of the diffusion in the buffer gas. The spin waves with high wave vector $|K|$ are washed off before the readout.

Apparently the mean values in Fig.~\ref{fig:avg} do not carry any information about   single-shot speckle pattern, coherence area or the intensity correlations of the Stokes and anti-Stokes. 
To extract and quantify interesting single-shot properties we calculate  statistical correlation of intensity fluctuations between portions of the scattered light $C_{ij}(\theta_x,\theta_y;\theta'_x,\theta'_y)$. 
It is defined using deviations of the actual photon number from the mean value $\Delta n_i(\theta_x,\theta_y)=n_i(\theta_x,\theta_y)-\avg{n_i(\theta_x,\theta_y)}$, $i=s,a$ as follows:
\begin{equation}
C_{ij}(\theta_x,\theta_y;\theta'_x,\theta'_y)
=\frac{\avg{\Delta n_i(\theta_x,\theta_y)\Delta n_j(\theta'_x,\theta'_y)}}
{\sqrt{\avg{\Delta n_i^2(\theta_x,\theta_y)}\avg{\Delta n_j^2(\theta'_x,\theta'_y)}}} 
\quad i=s,a\quad j=s,a
\end{equation}
We depict $C_{is}(\theta_x,\theta_y;\theta'_x,\theta'_y)$ with fixed reference direction $\theta'_x,\theta'_y$ in the Stokes as a function of $\theta_x,\theta_y$ for both the Stokes--Stokes $C_{ss}$ and anti-Stokes--Stokes $C_{as}$ type of correlations. 

\begin{figure}
\centering
\includegraphics[width=\textwidth]{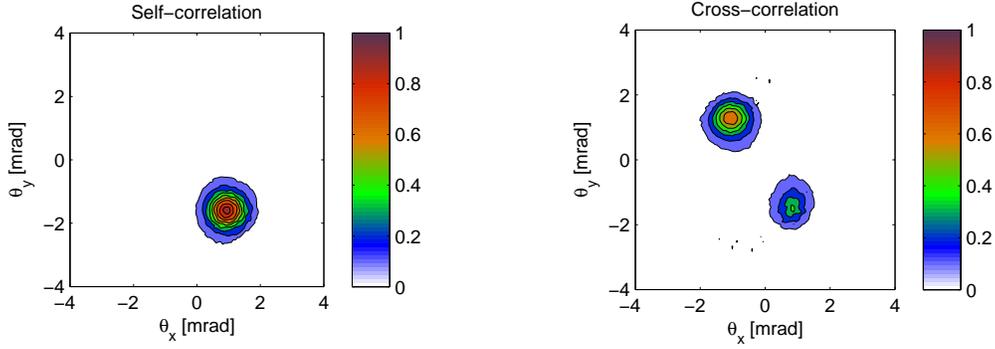}
\caption{Normalized intensity correlation functions $C_{ss}(\theta_x,\theta_y;\theta'_x,\theta'_y)$ (left panel) and $C_{as}(\theta_x,\theta_y;\theta'_x,\theta'_y)$ (right panel) between a fixed direction in the Stokes $\theta'_x,\theta'_y$ and any other direction $\theta_x,\theta_y$ in the Stokes and anti-Stokes respectively. 
The reference direction $(\theta'_x,\theta'_y)=(0.9,-1.6)$~mrad corresponds to the point in the center of the spot in the Stokes self-correlation function $C_{ss}$ in the left panel. 
The center of the best correlated spot corresponds to the conjugate direction of the  counterpart beam $(\theta_x,\theta_y)=-(\theta'_x,\theta'_y)$. 
Contours were drawn at levels of 0.1, 0.2 up to 0.9.
Results obtained in a cell with 1 torr Kr buffer gas.
}
\label{fig:corr}
\end{figure}

A typical correlations pictures obtained this way are shown in Fig.~\ref{fig:corr}. The spot in the Stokes self-correlation image $C_{ss}$ (left panel) represents the correlation between Stokes light emitted in similar directions, while the highest spot in the anti-Stokes cross-correlation $C_{as}$ (right panel in the figure) corresponds to a counter part correlated anti-Stokes light. The strength of the counter part beam correlations diminishes as we increase the distance from the driving beam, thus looking at larger $|K|$. Correlation pictures for other reference directions $\theta'_x,\theta'_y$ are very similar, the  conjugate spot always appears on the opposite side of the center with respect to the reference. 
Note the size of the correlation spot  is proportional to the size of a speckle in the single shot image Fig.~\ref{fig:single}  and it is larger by factor $\sqrt{2}$. This is explained by the fact that cross correlation value of each point is the sum of cross correlations of all the independent, displaced speckles. Therefore the actual cross-correlation function will be a convolution of the speckle shape with itself which for the Gaussian functions give the factor $\sqrt{2}$. 

Let us note that in the right, anti-Stokes--Stokes correlation panel Fig.~\ref{fig:corr} one can also see a second, weaker correlation spot.
It lies on the same side of the read beam as the reference point. This is a signature of further Stokes scattering induced by the read beam. It is possible because the read beam is detuned by approximately 5 GHz from the F=1$\,\rightarrow\,$F'=1 transition, which is apparently not enough to completely extinguish further amplification. 

\begin{figure}
\centering
\includegraphics[width=1\textwidth]{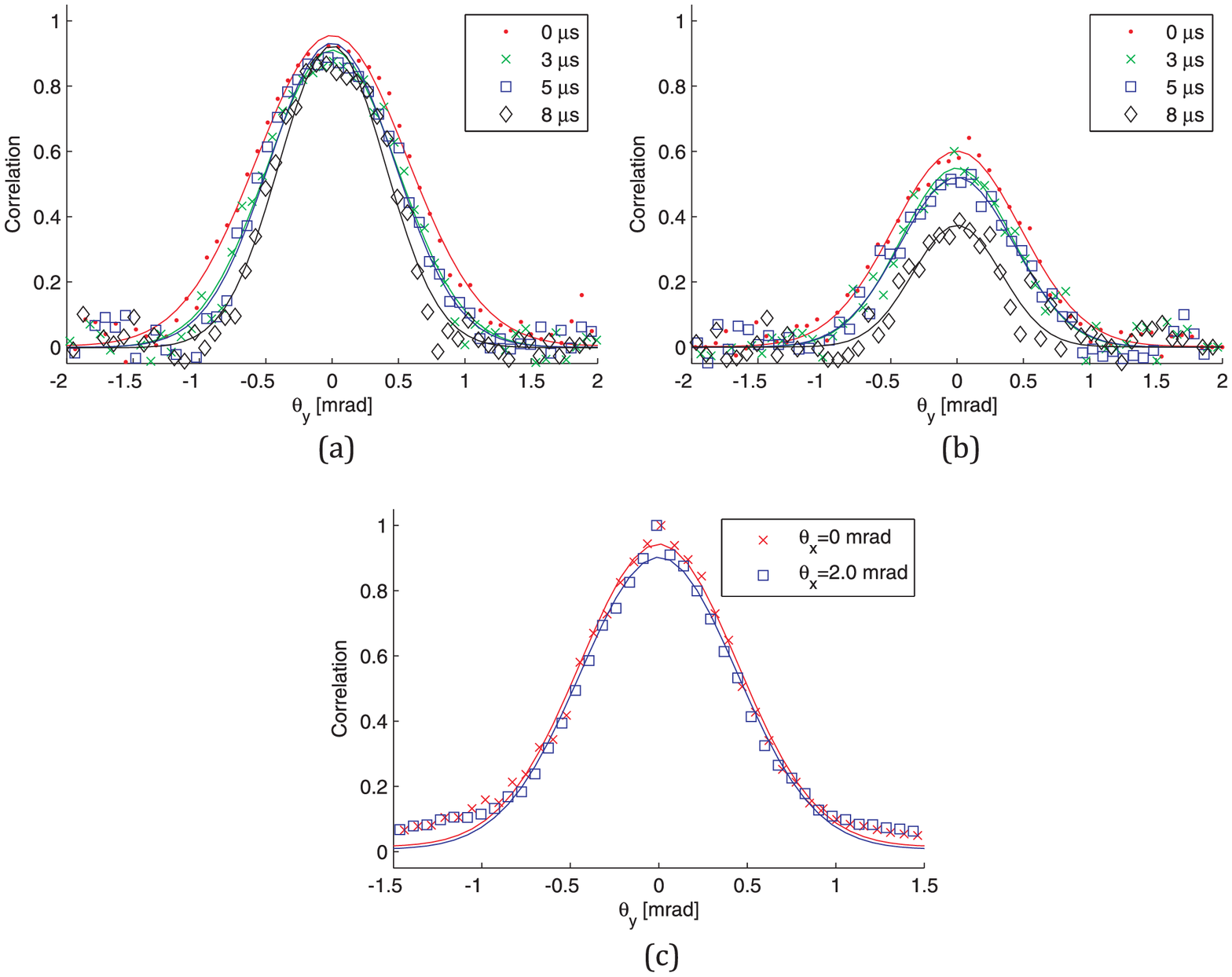}
\caption{Sections of the intensity correlation functions $C_{is}(\theta_x,\theta_y;\theta'_x,\theta'_y)$ along $\theta_y$. 
Cross-correlation $C_{as}$ for (a) $\theta'_x=\theta'_y=\theta_x=0$ and (b) $\theta'_x=-\theta_x=0.7$~mrad, $\theta'_y=0$ calculated for the subsequent storage times $t_s=$0, 3, 5, 8 $\mu$s.
(c) Self-correlation $C_{ss}$ for $\theta'_x=-\theta_x=0,\,2$~mrad, $\theta'_y=0$.
Points depict experimental data, lines correspond to Gaussian fits.}\label{fig:Ccs}
\end{figure}

Fig.~\ref{fig:Ccs} depicts cross sections through the correlation functions. 
In the first two panels Fig.~\ref{fig:Ccs}(a) and Fig.~\ref{fig:Ccs}(b) we compare the anti-Stokes--Stokes correlation functions $C_{as}$ for the reference direction along the write beam $\theta'_x=0$ and at an angle of $\theta'_x=0.7$~mrad to it. Initial width of the correlation function with respect to $\theta_y$ equals 1~mrad regardless of the reference direction $\theta'_x$. 
It shrinks first slowly and then quite abruptly for long storage times due to diffusional damping of high-$K$ spin waves.
In Fig.~\ref{fig:Ccs}(c) we can see correlation of the Stokes field with itself $C_{ss}$. It also has 1~mrad width regardless of the reference direction. 

The simple theory predicts that the Stokes field is composed of spatial modes \cite{Koodynski2012}. The number of modes $N$ can be approximated as a number of speckle spots that fit into the solid angle occupied by the scattered light. We take $N=2(w_\text{avg}/w_C)^2$, where $w_\text{avg}$ and $w_C$ are $1/e^2$ radius of the average intensity of the Scattered light and correlation function respectively. The numerical factor arises from the fact that the correlation function is square in the intensity. Its value is confirmed by considering a simple model, in which the registered intensity is a sum of randomly populated modes of a width $w_\text{speckle}$. It is found that the width of the correlation function is $w_C=\sqrt{2}w_\text{speckle}$. 
First we calculate the number of modes $N$ for the Stokes field. From the Gaussian fits we find the angular radius of the average intensity $w_\text{avg}=2.8$~mrad as depicted in Fig.~\ref{fig:avg} and Stokes-Stokes correlation function $w_C=1$~mrad, giving a total number of modes $N\simeq 15.7$.
These number allow us to asses the fundamental mode angular spread $w_0=w_\text{avg}/\sqrt[4]{N}=w_C\sqrt[4]{N/4}=1.4$~mrad which is of the order predicted by a simple theory $\Delta K=\sqrt{k_s/L}$ corresponding to 2.3~mrad. 

In the lossless case the number of modes $N$ is expected to be the Fresnel number of the write beam \cite{Koodynski2012}, $F=|k_s|w^2/(2\pi L)=15$ in our case, which is consistent with the experimental results. 
  Let us note that in experiments by Boyer \emph{et al.} \cite{Boyer2008b} they estimated the number of modes to be equal 100, which is also consistent with the Fresnel number calculated for their experimental conditions $F=120$. Therefore we conclude that the relation between number of modes and Fresnel number of the excited portion of the sample can be used for designing future experiments.
  
\begin{figure}
\centering
\includegraphics[scale=0.6]{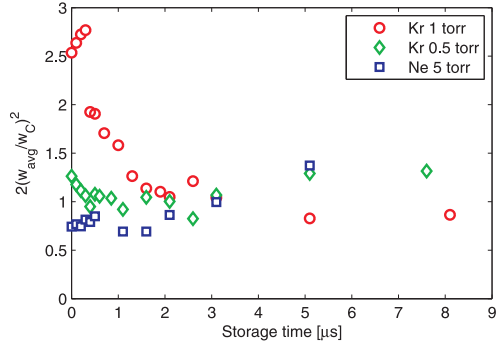}
\caption{The approximate number of modes in the anti-Stokes scattering $N=2(w_\text{avg}/w_C)^2$ as a function of storage time.}\label{fig:modes}
\end{figure}

In a similar manner we can calculate the number of retrieved modes as a function of time. For each storage time we calculate the average width of the anti-Stokes scattering and the correlation function. The result of estimated number of retrieved spatial modes is plotted in Fig.~\ref{fig:modes}.
We can see that only for 1 torr Krypton the number of retrieved modes in significantly greater than one, but still much lower than the number of stored modes we calculated. We suspect that this sudden drop results from a Doppler broadening in the atomic sample. The spin waves are imprinted on different velocity classes with varying phases \cite{Gorshkov2007c} that are inconsistent between write and read stages.
\section{Conclusion} \label{sec:summary}
In this article we characterize the performance of a  multimode delayed random image generator in which correlations are mediated between the beams via a spin wave excitation in the warm rubidium vapor. 
We turned our attention to the experimental characterization of the spatial behavior of the collective Raman scattering. 
In particular we quantified the influence of buffer gas pressure on the diffusion and pair production rate in the Stokes scattering. 
We measured the exponential decays of the spin waves to characterize the decoherence from atomic diffusion and we determined the diffusion constants for three different buffer gases. 
We compared these results to the angular dependence of the Stokes scattering growth parameters and found that it is dominated by the evolution of the fundamental most occupied mode.

We calculated the number of independent spatial modes excited in the generator. 
We showed how many can be retrieved after a certain storage time. 
Our results concerning the mode structure of scattered fields are consistent with a simple multimode model which permits space-time separation \cite{Koodynski2012}.

In future applications one may attempt to increase the number of modes. 
Our results suggest two ways to achieve that. 
One can apply the larger driving beam diameter. The number of modes,  determined by the Fresnel number, is expected to grow quadratically with a driving beam diameter if constant laser intensity is applied. An increase of a beam diameter would also shorten the wave vectors of the excited spin waves, making them more resilient to diffusion. 
The other way to increase the number of modes is to slow down the atomic diffusion keeping a collisional decoherence low. 
Our results indicate a need for further comparisons of various buffer gases to find an optimum. 
This type of experiment could be also implemented in cold atomic gases and possibly in doped crystals.

 In this work we studied only classical properties of the light generated. 
However, based on a number of similar successful experiments \cite{Marino2008,Boyer2008b} we conjecture the conclusions on the spatial properties of the light would stay valid also in the low light quantum regime. Our results could therefore be of use for a number of emerging multimode quantum technologies. 
In particular the quantum information processing protocols devised for warm atomic samples could benefit from multimode capacity of the storage, because a single cell operated in multimode regime is equivalent to multiple single mode cells. 
\section*{Acknowledgements} 
We acknowledge fruitful discussions with Thomas Fernholz and Czes{\l}aw Radzewicz as well as generous support from Konrad Banaszek.
This work was supported by the Foundation for Polish Science TEAM project co-financed by the EU European Regional Development Fund and FP7 FET project Q-ESSENCE (Contract No. 248095). W.W. was also supported by the National Science Centre grant no. DEC-2011/03/D/ST2/01941.
\end{document}